\documentclass[aps,prl,amsmath,amssymb,reprint,showpacs,superscriptaddress,citeautoscript,twocolumn]{revtex4}
\usepackage{helvet}
\usepackage{mathptmx}
\usepackage[utf8]{inputenc} 
\usepackage{graphicx}
\usepackage{xspace}
\usepackage{epstopdf}
\usepackage[
,letterpaper
,textwidth=17.0cm
,textheight=22.3cm
,verbose
,dvips
]{geometry}

\begin{document}

\title{Epitaxial interfaces between crystallographically mismatched materials}
\author{Steven C. Erwin}
\affiliation{Center for Computational Materials Science, Naval Research Laboratory, Washington, DC 20375}
\author{Cunxu Gao}
\author{Claudia Roder}
\author{Jonas Lähnemann}
\author{Oliver Brandt}
\affiliation{Paul-Drude-Institut für Festkörperelektronik, Hausvogteiplatz 5--7, 10117 Berlin, Germany}
\date{\today}

\begin{abstract}
  We report an unexpected mechanism by which an epitaxial interface
  can form between materials having strongly mismatched
  lattice constants. A simple model is proposed in which one material
  tilts out of the interface plane to create a coincidence-site
  lattice that balances two competing geometrical criteria---low
  residual strain and short coincidence-lattice period. We apply this
  model, along with complementary first-principles total-energy
  calculations, to the interface formed by molecular-beam epitaxy of
  cubic Fe on hexagonal GaN and find excellent agreement between
  theory and experiment.
\end{abstract}


\maketitle

A fundamental goal of materials science is to elucidate and exploit
the physical principles that govern epitaxial growth
\cite{bauer_j_mater_res_1990a,palmstrom_ann_rev_mat_sci_1995a}. Some
of these principles are well-established. For example, if the lattice
constants of the film and substrate are close but not identical then
a coherently strained film may grow up to a critical thickness, beyond
which misfit dislocations relieve the strain
\cite{matthews_j_crystal_growth_1974a}. Alternatively, a film and substrate having
lattice spacings close to an integer ratio $m/n$ may form an
epitaxial interface described by a coincidence lattice 
\cite{sutton_acta_metall_1987a,trampert_physica_e_low_dimen_sys_nanostr_2002a}.

In this Letter we report a new and unexpected mechanism by which
epitaxial films can grow on substrates having, in principle, an {\em
  arbitrary} lattice mismatch. We illustrate this mechanism
experimentally by growing single-crystal Fe on $M$-plane GaN. The Fe
grows in an unusual crystallographic orientation with a very high
Miller index, Fe(205).  We develop a simple theoretical model which,
when complemented with total-energy calculations, correctly predicts
this exact orientation as well as the single-domain nature of the
film.  Finally, we use our model to propose a new strategy for growing
nonpolar epitaxial GaN films on high-index Si substrates.

% FIG tem
\begin{figure} 
\includegraphics[width=8cm]{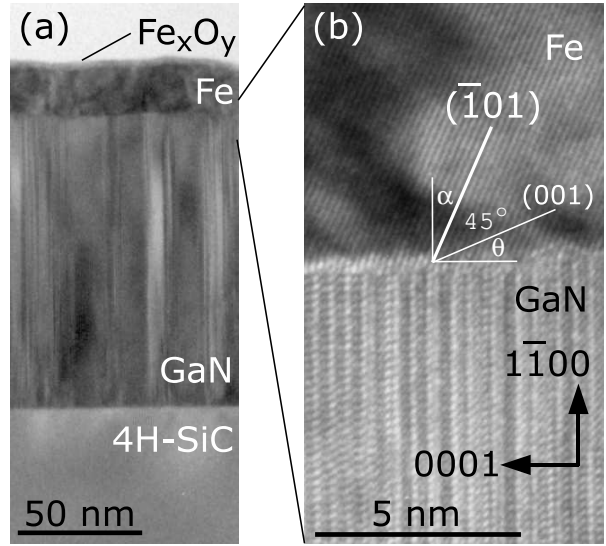} 
\caption{(a) Cross-sectional transmission electron micrograph taken
  along the $[\overline{1}\overline{1}20]$ zone axis of the Fe/GaN/4H-SiC
  structure under investigation. 
  The top 2~nm of the Fe film are oxidized. (b) Cross-sectional
  high-resolution transmission electron micrograph of the
  Fe/GaN$(1\overline{1}00)$ interface along the
  $[\overline{1}\overline{1}20]$ zone axis. The Fe($\overline{1}$01)
  and GaN(0001) lattice planes form an
  angle, $\alpha$, 
  of approximately $23^\circ$.  Stacking faults in the GaN, visible in both panels,
  do not affect the orientation of the Fe film.}
\label{tem}
\end{figure}

The epitaxial growth of both GaN and Fe was performed in a
custom-built molecular-beam epitaxy system equipped with solid-source
effusion cells for Ga and Fe. Active nitrogen was provided by a
radio-frequency N$_2$ plasma source. Nucleation and growth were
monitored \emph{in situ} by reflection high-energy electron
diffraction. A 130-nm thick layer of $M$-plane GaN was first grown on
a 4H-SiC$(1\overline{1}00)$ substrate under Ga-stable conditions and a
temperature of 720$^\circ$C. After growth of the GaN layer, excess Ga
was desorbed prior to cooling down to 350$^\circ$C for the deposition
of Fe \cite{brandt_phys_rev_b_2004a}. The Fe film grew at this temperature at a rate of 0.13~nm/min
to a final thickness of 27~nm. The resulting Fe/GaN/SiC
heterostructure was investigated by cross-sectional transmission
electron microscopy (TEM) and convergent-beam electron diffraction
(CBED) using a JEOL JEM-3010 operating at 300~keV. Electron
backscattering diffraction (EBSD) was carried out in a Zeiss Ultra-55
scanning electron microscope equipped with an EDAX-TSL EBSD system.

% FIG schematic
\begin{figure} 
\includegraphics[width=7cm]{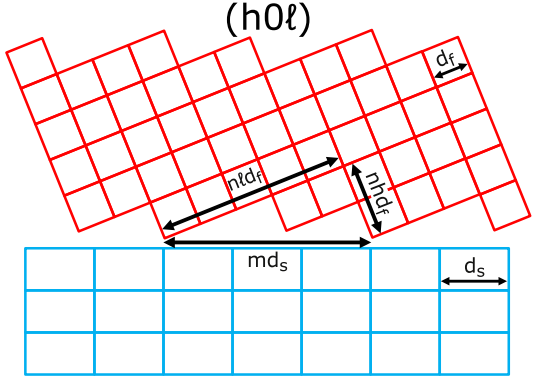} 
\caption{Schematic view of a commensurate interface between a film
  with lattice constant $d_f$ and a substrate with lattice constant
  $d_s$.  The interface plane is $(h0 \ell)$ with respect to the film,
  corresponding to a tilting from (001) toward (101) by the angle
  $\theta = \tan^{-1}(h/\ell)$. The condition for
  commensurability is that a coincidence-site lattice, defined by a
  pair of integers $(m,n)$, exists such that $md_s =
  n(h^2+\ell^2)^{1/2}d_f$. The example depicted here is (205) with
  $\theta = 21.8^\circ$ (equivalent to the angle $\alpha=
  23.2^\circ$ as defined in Fig.~\ref{tem}) and
  $(m,n)=(3,1)$.}
\label{schematic} 
\end{figure}

Figure~\ref{tem}(a) shows a cross-sectional transmission electron
micrograph of the Fe/GaN/4H-SiC structure. Despite the structural and
chemical differences of the constituent materials, the respective
layers are well defined and exhibit abrupt interfaces. The GaN layer
is seen to contain stacking faults due to the stacking mismatch
between 2H-GaN and 4H-SiC. Nevertheless, the high-resolution detail of the Fe/GaN
interfacial region shown in Fig.~\ref{tem}(b) reveals an epitaxial
relationship between the Fe film and the underlying
GaN$(1\overline{1}00)$ layer. The ($\overline{1}$01) lattice planes of
the Fe film are clearly resolved and are found to be well ordered,
unaffected by the stacking disorder in the GaN layer. Of
special interest is the angle, approximately 23$^\circ$, formed by the
Fe ($\overline{1}$01) planes and the vertical interface normal. 
This angle indicates that the Fe interface plane has a high Miller
index---an unexpected finding in light of the comparatively high
surface energies of high-index metal surfaces.  
We show now that precisely this orientation is predicted by a
simple, physically transparent model (complemented with first-principles
total-energy calculations) of epitaxial interfaces between
dissimilar materials.

Consider the formation of an interface between a film ($f$) and a
substrate ($s)$ having different lattice constants $d_f$ and $d_s$. If
the lattice mismatch is sufficiently small then the strained film may
grow coherently until it reaches its critical thickness
\cite{matthews_j_crystal_growth_1974a}. For much larger mismatch this
scenario becomes unlikely.  Epitaxial growth is nevertheless possible
by tilting the orientation of the film, as the TEM image in
Fig.~\ref{tem}(b) makes clear. Figure ~\ref{schematic} illustrates how
an arbitrary lattice mismatch can be accommodated by allowing the film
to have an orientation between (001) and (101) given by the Miller
indices $(h0 \ell)$.  Our goal below is to predict the most stable
film orientation given the lattice constants $d_f$ and $d_s$.

The unit cell of a film with orientation $(h0 \ell)$ has length
$L_f=(h^2+\ell^2)^{1/2}d_f$. In order for the film and substrate to be
commensurate there must exist a coincidence-site lattice (CSL),
defined by a pair of integers $(m,n)$, such that $md_s = n L_f$. This
condition is unrealistically restrictive, however.
In real
systems the film will tolerate a small compressive or tensile strain
$\varepsilon_{xx}$ which relaxes the CSL condition to $md_s = n
L_f(1+\varepsilon_{xx})$.

% FIG fireworks
\begin{figure} 
\includegraphics[width=8cm]{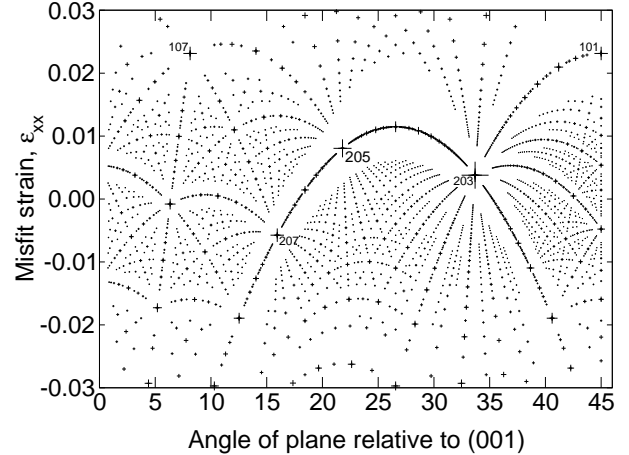} 
\caption{Misfit strain in commensurate interfaces between bcc Fe and
  GaN$(1\overline{1}00)$. The points represent all possible Fe$(h0
  \ell)$ planes having $h\le \ell\le75$.  The misfit
  $\varepsilon_{xx}$ is the strain component along the [0001]
  direction of the GaN.  The size of each plot symbol is inversely
  proportional to the period $m$ of the coincidence-site lattice (CSL)
  that minimizes the misfit. Fe$(h0 \ell)$ planes having both small
  misfit strain and small CSL period are labeled.} \label{fireworks}
\end{figure}

We propose two geometrical criteria for identifying candidate
orientations for interfaces with low energy. (1) The misfit strain
$\varepsilon_{xx}$ should be as small as possible, and (2) the period
$m$ of the CSL should be as small as possible. The latter criterion is
motivated by analogy to low-energy grain boundaries between two
identical materials, which often have a CSL with small unit cell
volume $\Sigma$
\cite{sutton_acta_metall_1987a}.
For interfaces
between different materials it is not generally possible to minimize
the strain and CSL period simultaneously. Nor is it obvious how to
construct a single objective function of both which could then be
optimized.  Instead, we apply both criteria with the aim of selecting
a small subset of candidate orientations for subsequent study with a
more quantitative method such as density-functional theory (DFT).

To apply these criteria to the growth of Fe on the $M$-plane of GaN we
equate $d_f$ with the bcc Fe lattice constant, 2.866
\AA~\cite{mao_j_appl_phys_1967a}, and $d_s$ with the GaN $c$ lattice
parameter, 5.186 \AA~\cite{leszczynski_appl_phys_lett_1996a}. Figure
\ref{fireworks} shows the resulting Fe misfit strain needed to satisfy
the CSL condition for a large number of hypothetical orientations of
the Fe film.  In this plot each orientation $(h0 \ell)$ is represented
by its angle $\theta=\tan^{-1}(h/\ell)$ relative to the (001)
plane. The period of each CSL is encoded by the size of the plot
symbol, which is inversely proportional to $m$. Only points with small
strains, less than 3\%, are displayed here. 

% FIG model
\begin{figure} 
\includegraphics[width=8.0cm]{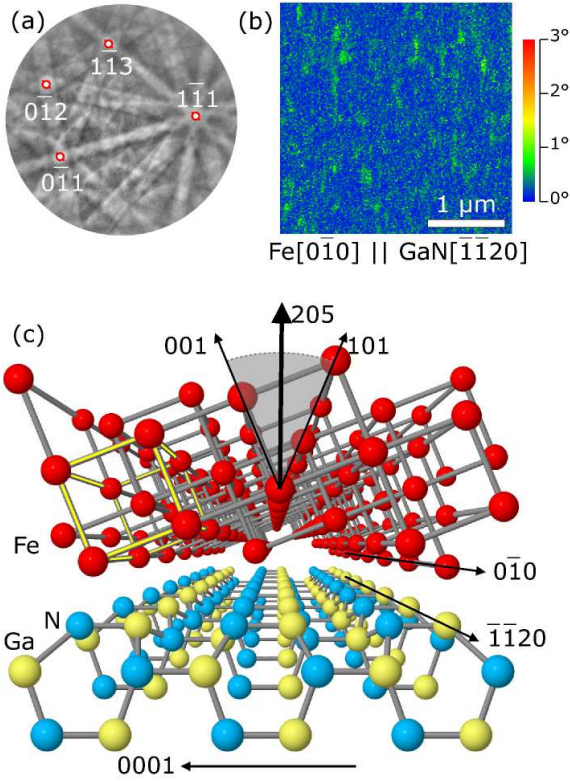} 
\caption{(a) Pseudo-Kikuchi pattern of the Fe film on
  GaN$(1\overline{1}00)$ depicted in Fig.~\ref{tem}.  The major
  low-index zone axes are indicated. (b) In-plane orientation map of
  the Fe film. Variations in color encode local deviations from the
  nominal orientation.  The film has the same orientation over the
  entire area within a tolerance of 1$^{\circ}$. (c) Model of the
  Fe(205)/GaN$(1\overline{1}00)$ epitaxial interface. For clarity an
  ideal geometry with arbitrary registry is shown here. Slightly less
  than one unit cell along [0001] is depicted. A conventional unit
  cell of Fe is highlighted in yellow.  }
\label{model} 
\end{figure}

The rich structure visible in Fig.~\ref{fireworks} makes it clear that
the misfit strain can be made arbitrarily small for many different
film orientations. Hence the strain alone cannot provide a definitive
criterion favoring a particular growth plane. Moreover, the vast
majority of these low-strain orientations require a very large CSL
period and hence do not constitute physically meaningful
commensurability. Only a very few orientations offer both a small
strain and small CSL period, namely (207), (205), and (203). The (205)
orientation corresponds to the angle $\theta= 21.8^\circ$. This
is equivalent to $\alpha=45-\theta=23.2^\circ$ as defined in
Fig.~\ref{tem}(b) and thus is in excellent agreement with the measured
angle, 23$^\circ$, obtained from TEM.

It is important to realize that our purely geometrical criteria do
not distinguish between the orientation (205) depicted in
Fig.~\ref{schematic} and its symmetry-equivalent counterpart,
(20$\overline{5}$) ($\theta=-21.8^\circ$), created by rotating the
film by 180$^\circ$ about the substrate normal. Indeed, the
pairs [$(h0\ell)$, $(h0\overline{\ell})$] have the same strain and
CSL and hence are equivalent within this model. 
In the Fe/GaN system, however,
the polarity of the GaN wurtzite structure breaks this equivalence. The
question that then arises is whether the influence of the polarity is sufficiently
strong to select a single orientation, and if so, which one?
The TEM image in Fig.~\ref{tem} shows a single 
orientation but is limited to a nanometer-scale region of the film.
To characterize a much larger area we used EBSD
\cite{nolze_cryst_res_technol_2005a}.  Figure~\ref{model}(a) shows the
resulting pseudo-Kikuchi pattern of the Fe film in Fig.~\ref{tem}. The
pattern exhibits sharp and well defined Kikuchi bands, reflecting high
crystal quality of the Fe film and allowing for a fast and reliable
indexing of the patterns recorded while scanning the electron beam
over a large area (9~$\mu$m$^2$). The resulting EBSD map shown in
Fig.~\ref{model}(b) visualizes the in-plane orientation of the Fe film
with a spatial and angular resolution of 20~nm and 1$^\circ$,
respectively. The map reveals the complete absence of any domain
structure. Indeed, the film is single crystalline, and has the same
orientation as found by TEM throughout the mapped area.

To understand why a single orientational domain is found requires
going beyond a model based solely on interface geometry. Now the
interface structure---its precise atomic arrangement and chemical
bonding---must be addressed.  To do this we used 
DFT to calculate the relative formation energies of finite Fe
films grown on the $M$-plane of GaN. We considered the six different
orientations predicted by the geometrical model to be favorable:
(207), (205), (203), and their rotated counterparts
(20$\overline{7}$), (20$\overline{5}$), (20$\overline{3}$). The
different films were the same thickness, 6 \AA, equivalent to about
four monolayers.  Figure \ref{model}(c) depicts the
Fe(205)/GaN$(1\overline{1}00)$ interface as an example.

We have previously shown that at 3--4 monolayers the contribution of
the Fe/GaN interface formation energy to the full formation energy of
the film is already converged \cite{gao_phys_rev_b_2010a}. We also
find that the Fe free surface energy varies by less than 1 meV/\AA$^2$
among the three orientations we consider here
\cite{erwin_unpublished}. Therefore the relative formation energy of
the finite film closely mirrors, with good accuracy, the formation
energy of the isolated interface.

For each orientation the Fe film was slightly strained along the $x$
[GaN(0001)] direction according to its CSL as discussed above. There
is also a lattice mismatch in the $y$
[GaN$(\overline{1}\overline{1}20)$)] direction because the Fe lattice
constant and GaN $a$ lattice parameter differ by nearly 12\%.  This
mismatch was accommodated by a single CSL, common to all orientations,
containing eight unit cells of Fe and seven of GaN. The GaN substrate
was represented by a slab of four atomic layers with fixed in-plane
equilibrium lattice parameters and a passivating bottom layer.
Total energies and forces were calculated within the PBE generalized-gradient approximation
\cite{perdew_phys_rev_lett_1996a} to DFT
using projector-augmented-wave potentials as implemented in {\sc
vasp} \cite{kresse_phys_rev_b_1993a,kresse_phys_rev_b_1996a}.
All Fe and GaN atomic positions were fully relaxed except the bottom
GaN layer. For each Fe film orientation the formation energy depends
strongly on the choice of Fe-GaN interface registry. We systematically
varied the registry over a grid in both $x$ and $y$ to locate the
global energy minimum for each orientation.  The plane-wave cutoff for
all calculations was 400~eV.

The resulting formation energies are listed in Table~\ref{energies}.
The most favorable orientation is Fe(205). Of the candidates tested,
this orientation has neither the smallest possible strain nor the
shortest possible CSL period, indicating that the optimal interface
structure is an important third criterion that must supplement the two
geometrical criteria. Note also that the small variation in the
formation energy of the Fe surface, which is included in the formation energy of the film,
is too small to affect the overall energy ordering of the
orientations.

% TABLE
\begin{table}
  \caption{Formation energies calculated within density-functional theory of Fe/GaN$(1\overline{1}00)$ films for the Fe planes in Fig.~\ref{fireworks} that have both a small misfit strain and a coincidence-site lattice (CSL) with small period $m$.  The Fe film thickness is the same for all cases, approximately 6 \AA.  Formation energies are relative to the most favorable plane, Fe(205), in units of meV/\AA$^2$.} 
\begin{ruledtabular}

\begin{tabular}{cccc} 
\textrm{Plane}& \textrm{Misfit strain}& \textrm{CSL period}& \textrm{\mbox{   }Formation energy}\\ 
\colrule 
Fe($205$) &  $+0.008 $ & 3 &   0  \\ 
Fe($20\overline{5}$) &  $+0.008 $ & 3 &   4  \\ \\
Fe($203$) &  $+0.004 $ & 2 &   9  \\ 
Fe($20\overline{3}$) &  $+0.004 $ & 2 &  16  \\ \\ 
Fe($207$) &  $-0.006 $ & 4 &  12  \\ 
Fe($20\overline{7}$) &  $-0.006 $ & 4 &  13  \\ 
\end{tabular} 
\end{ruledtabular} 
\label{energies}
\end{table}

To distinguish Fe(205) from Fe(20$\overline{5}$) experimentally
requires determining the absolute polarity of the GaN substrate.
We did this by recording CBED patterns
with a beam spot size of approximately 15~nm under two-beam
conditions. Simulations of CBED patterns were performed using {\sc jems}
\cite{stadelmann_jems_1999a} to index the crystallographic directions
observed in the experimental patterns and thus to determine the
polarity.  The resulting absolute orientation relationship is
Fe$(205)[0\overline{1}0]\vert\vert$GaN$(1\overline{1}00)[\overline{1}\overline{1}20]$,
in agreement with the prediction of our theoretical model complemented
with the results of DFT calculations.

Our model suggests a new strategy for growing non-polar GaN films.
The basic idea is to turn
Fig.~\ref{schematic} upside-down and consider the growth of $M$-plane
GaN on a suitable high-index substrate.  For a given 
material this requires identifying candidate orientations 
corresponding to small strain and small CSL period. One very promising
material is Si, which is already in widespread use as a
flat substrate for GaN/Si epitaxy despite the resulting high
dislocation densities \cite{joblot_superlatt_microstr_2006a}. Many
high-index Si substrates are readily available, and some have already
been used for GaN growths \cite{ni_appl_phys_lett_2009a,ravash_appl_phys_lett_2010a}.
Calculations are in progress to identify promising high-index Si 
orientations for growing nonpolar GaN with low strain
\cite{kutana_unpub_2011a}.

%\bibliography{database} 

\end{document}